\documentclass{moriond}
\pdfoutput=1

\usepackage{graphicx}
\usepackage{color}

\def\be{\begin{equation}}
\def\ee{\end{equation}}
\def\bea{\begin{eqnarray}}
\def\eea{\end{eqnarray}}

\newcommand{\UNNLOPS}{\protect\scalebox{0.9}{UN$^2$LOPS}\xspace}
\newcommand{\UNLOPS}{\protect\scalebox{0.9}{UNLOPS}\xspace}

\newcommand{\POWHEG}{\protect\scalebox{0.9}{POWHEG}\xspace}

\newcommand{\MCatNLO}{\protect\scalebox{0.9}{MC@NLO}\xspace}

\newcommand{\Sherpa}{\protect\scalebox{0.9}{SHERPA}\xspace}

\usepackage{amsmath,amssymb,graphicx,xspace,subfigure,relsize,todonotes}

\def\halfplotwidth{0.35}

\newcommand{\Photo}{}

\usepackage{eso-pic}

\begin{document}

\AddToShipoutPictureBG*{%
  \AtPageUpperLeft{%
    \hspace{\paperwidth}%
    \raisebox{-3\baselineskip}{%
      \makebox[0pt][r]{SLAC-PUB 16334\hspace{2cm}}
}}}%

\vspace*{1cm}
\title{Combining parton showers and NNLO matrix elements}

\author{ S. H\"{o}che, Y. Li, S. Prestel\footnote{Speaker}}

\address{
Theory group,
SLAC National Accelerator Laboratory,\\
Menlo Park, CA 94025, USA}

\maketitle\abstracts{
In this talk, we discuss recent developments in combining parton showers
and fixed-order calculations. We focus on the \UNNLOPS method for matching
next-to-next-to-leading order computations to the parton shower, and we
present results from \Sherpa for Drell-Yan lepton-pair and Higgs-boson
production at the LHC.
}

\section{Introduction}
\label{sec:introduction}

With the LHC experiments entering the second long phase of data collection after
the upgrade period, we expect that the Standard Model (SM) of particle physics
will be probed in exquisite detail while searching for hints of phenomena beyond
our current knowledge. A major role in this endeavor is played by parton-shower
Monte Carlo programs, which allow to predict the full final-state kinematics on an
event-by-event basis.

In this talk, we will briefly describe the evolution and status of combining
fixed-order calculations with parton shower (PS) resummation, followed by
comments on which state-of-the art merging schemes lend themselves to further
improvements. We will then discuss how next-to-next-to-leading order (NNLO) 
accurate predictions can be included into event generators. Finally, we present 
results in the \UNNLOPS scheme~\cite{Hoeche:2014aia,Hoche:2014dla} as 
implemented in \Sherpa~\cite{Gleisberg:2008ta}. 

%

\section{The story so far}
\label{sec:story_so_far}

Finding ways to combine accurate fixed-order calculations with parton showers
has been a major topic in event generator development since the turn
of the century. A decisive boost came from methods for merging multiple 
inclusive tree-level calculations by making them exclusive using
Sudakov form factors derived from the parton shower~\cite{Catani:2001cc,Lavesson:2008ah,Lonnblad:2012ix}.
Another breakthrough was the development of algorithms for matching parton showers to 
NLO QCD calculations~\cite{Frixione:2002ik}.

All these methods have ambiguities and uncertainties. A particularly striking 
example of differences between NLO+PS matched results was presented in~\cite{Alioli:2008tz}: 
The prediction for the Higgs-boson transverse momentum distribution shown in this publication
varies greatly with the matching scheme. Differences in the schemes are formally 
beyond the required NLO+PS accuracy. Their numerical size reveals, however, 
that more accurate and less variable calculations of the Higgs-boson + jet process
must be included to make experimentally relevant predictions. 

This can be achieved using methods for combining a sequence of multi-parton 
fixed-order calculations, often referred to as "multi-jet merging".
Merging methods exist for tree-level~\cite{Catani:2001cc}
 and NLO calculations~\cite{Lavesson:2008ah,Lonnblad:2012ix}.
They provide state-of-the art predictions for LHC Run-II. 
A comparison of NLO merging schemes in~\cite{Butterworth:2014efa} 
has shown good agreement between different approaches.
More importantly, the agreement between theory and experiment is improved, 
and theoretical uncertainties may be reduced.

%

\section{Moving towards NNLO accuracy}
\label{sec:towards_nnlo_matching}

NLO multi-jet merging techniques have additional features compared to
LO merging. For example, those real-emission corrections to 
$X+n$-jet production which lead to $n+1$ well-separated jets above the 
merging scale need to be removed, since such configurations are
already included by merging with the $n+1$-jet calculation. 
In addition, the approximate virtual corrections included in the PS
must, at $\mathcal{O}(\alpha_s^{n+1})$, be replaced by the full NLO result.
A more subtle issue arises from additionally demanding the stability of 
inclusive jet cross sections~\cite{Lonnblad:2012ix,Hamilton:2012rf}:
In merged calculations, the emission probability is given by exact
fixed-order matrix elements. In contrast, the resummed virtual corrections
derive from the Sudakov factor of the parton shower. Upon integration
over the radiative phase space, the two do not cancel, leading to a 
``unitarity violation''. 

This discrepancy can be removed using unitary merging techniques~\cite{Lonnblad:2012ix}. 
One of them is the so-called \UNLOPS method. It allows, in a process-independent way,
to add the precise difference between fixed-order real-emission matrix elements and 
their parton-shower approximations to the merged result. This is called the 
"subtract what you add" philosophy. In the \UNLOPS scheme, it is possible to 
combine arbitrarily many NLO calculations, and include tree-level results 
when NLO calculations are not available. \UNLOPS retains the merging scale as a 
\emph{technical} parameter, since low merging scales  -- while desirable to use 
higher-order calculations over most of the phase space -- 
leads to inefficient event generation.

\section{Combining NNLO calculations with parton showers}
\label{sec:nnlo_matching}

Although NLO merging yields accurate predictions for many multi-jet observables,
it is desirable for some reactions to move beyond NLO accuracy. Such processes
include reactions with large higher-order corrections, e.g.\ Higgs-boson 
production in gluon fusion, standard candles like Drell-Yan lepton pair 
production, and other phenomenologically important processes.

NNLO accurate matching to the parton shower has been achieved first in the MINLO approach~\cite{Hamilton:2013fea}. 
The MINLO method~\cite{Hamilton:2012rf} is based on matching the hard process plus one-jet NLO calculation 
to the parton shower, and supplement it with Sudakov form factors that account for the resummed virtual
and unresolved higher-order corrections between the hard scale and the resolution scale of the jet.
In its current implementation it uses analytic Sudakov factors derived for $q_T$ resummation, which limits
its applicability to hard processes with no light QCD jets in the final state.
The genuine NNLO corrections are included through pre-tabulated phase-space dependent 
K-factors, which leads to fast event generation but makes the extension to
processes with more complicated final states challenging.

Within the \UNNLOPS approach~\cite{Hoeche:2014aia,Hoche:2014dla}, a variant of \UNLOPS,
NNLO corrections associated with the emission of resolvable
QCD radiation are treated as the hard process plus one additional jet at NLO. The remainder
of the phase space is filled by a calculation for the hard process at NNLO, with a corresponding
veto on any QCD activity. Both parts are separately finite, and parton shower matching is only
needed for the first. To make the result physically meaningful, the separation cut must be smaller
than the infrared cutoff of the parton shower. This requires very stable NLO matched calculations
for the one-jet process. In contrast to the MINLO method, real-emission configurations do not 
receive a contribution from the NNLO K-factor. 

Neither NNLOPS nor \UNNLOPS should be considered final a answer to NNLO+PS matching, 
but rather as a first step towards more general methods.

\section{NNLO+PS matched results in \Sherpa}
\label{sec:results}

We will now discuss some phenomenologically relevant results obtained with the \UNNLOPS 
matching as implemented in the \Sherpa event generator. In order to control all 
aspects of the matched calculation, the full NNLO calculation using a $q_\perp$ cutoff method
has been implemented in Sherpa itself. This technique is limited to processes without light jets
in the hard process, a shortcoming that can in principle be remedied by using different 
techniques for performing the fixed-order NNLO calculation. The following plots, and the 
\Sherpa plug-in containing the \UNNLOPS implementation are publicly available~\cite{Code}.

\begin{figure*}[t]
\centering
\includegraphics[width=\halfplotwidth\textwidth]{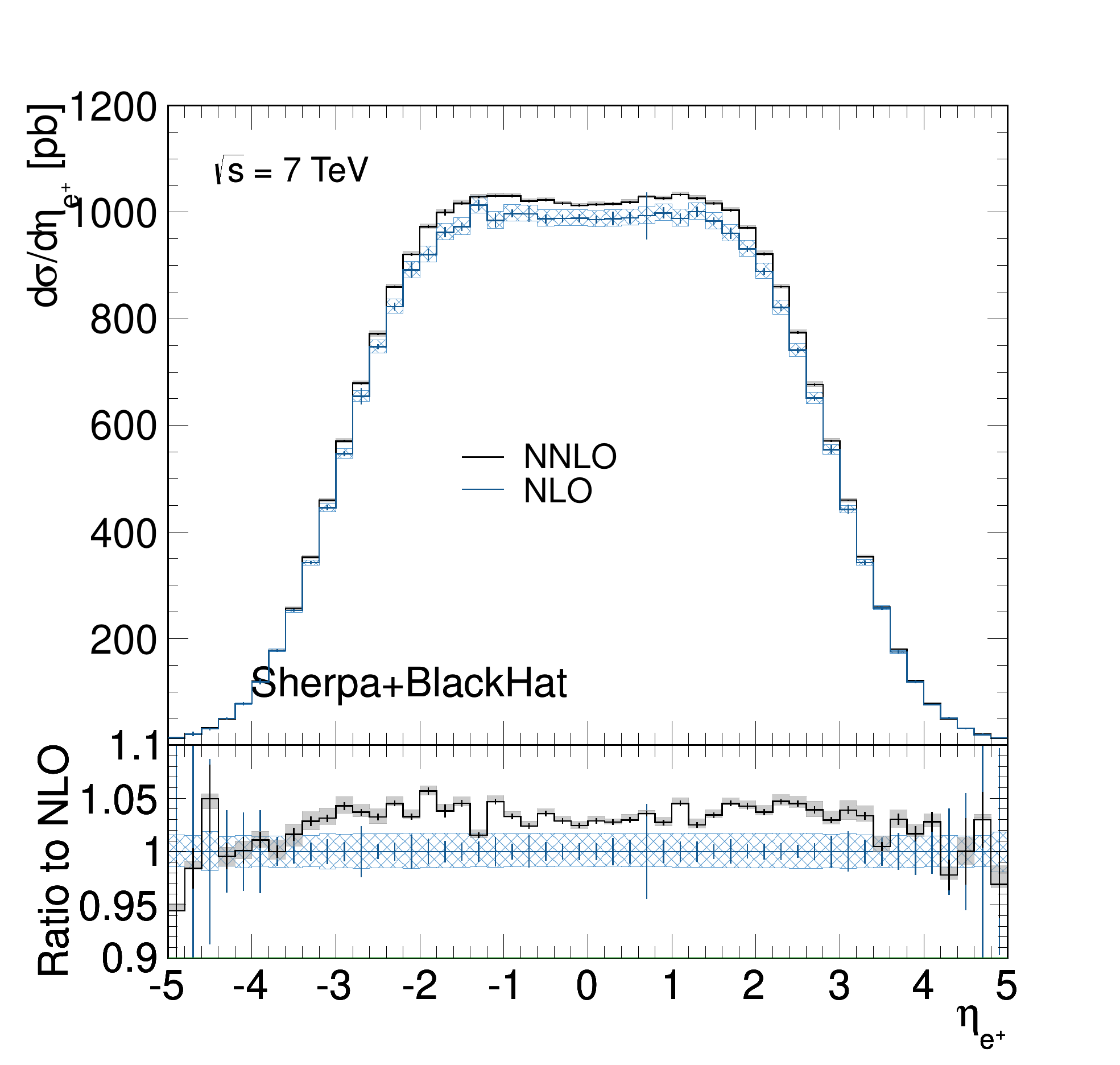}
\includegraphics[width=\halfplotwidth\textwidth]{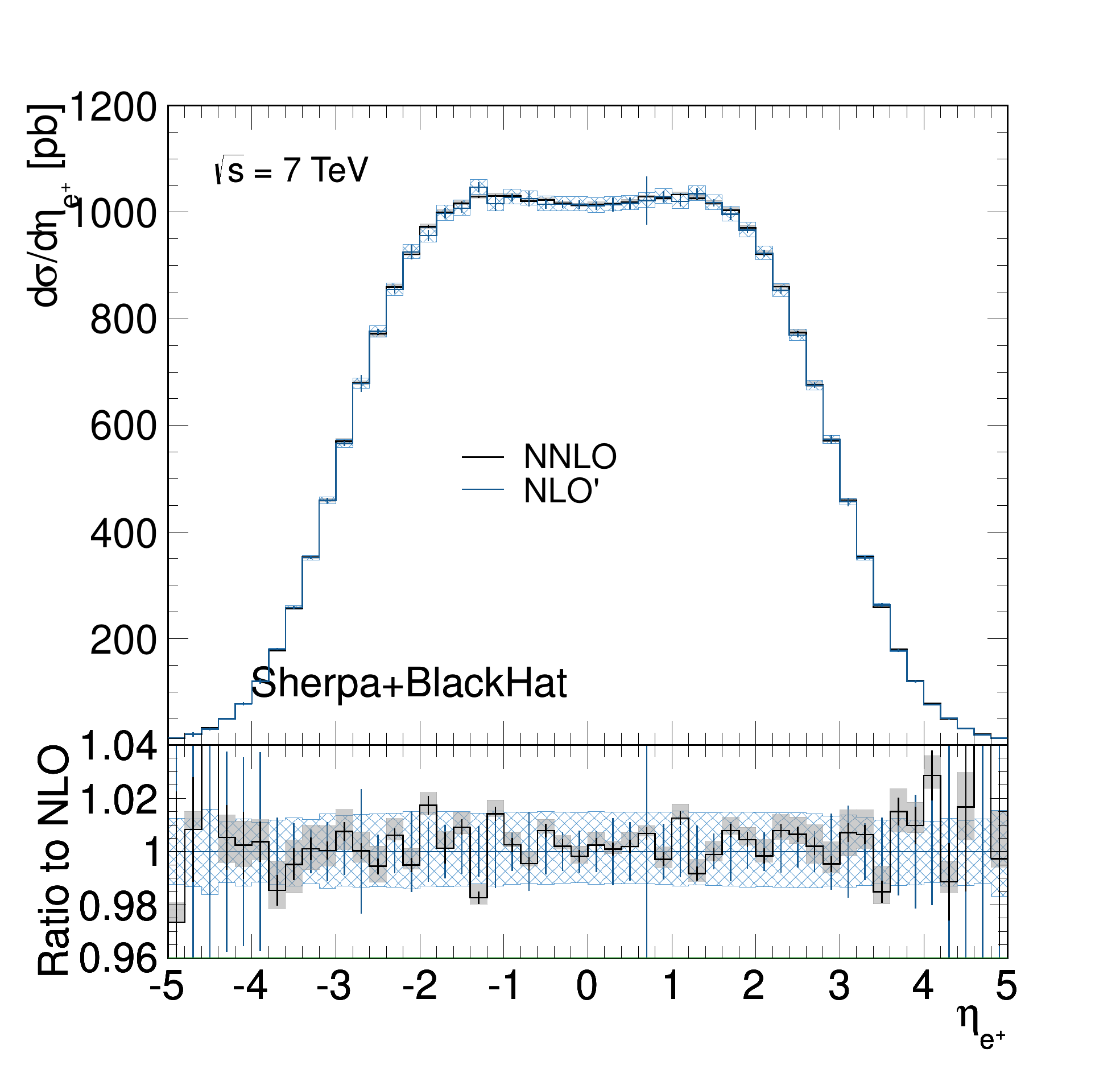}\\
\includegraphics[width=\halfplotwidth\textwidth]{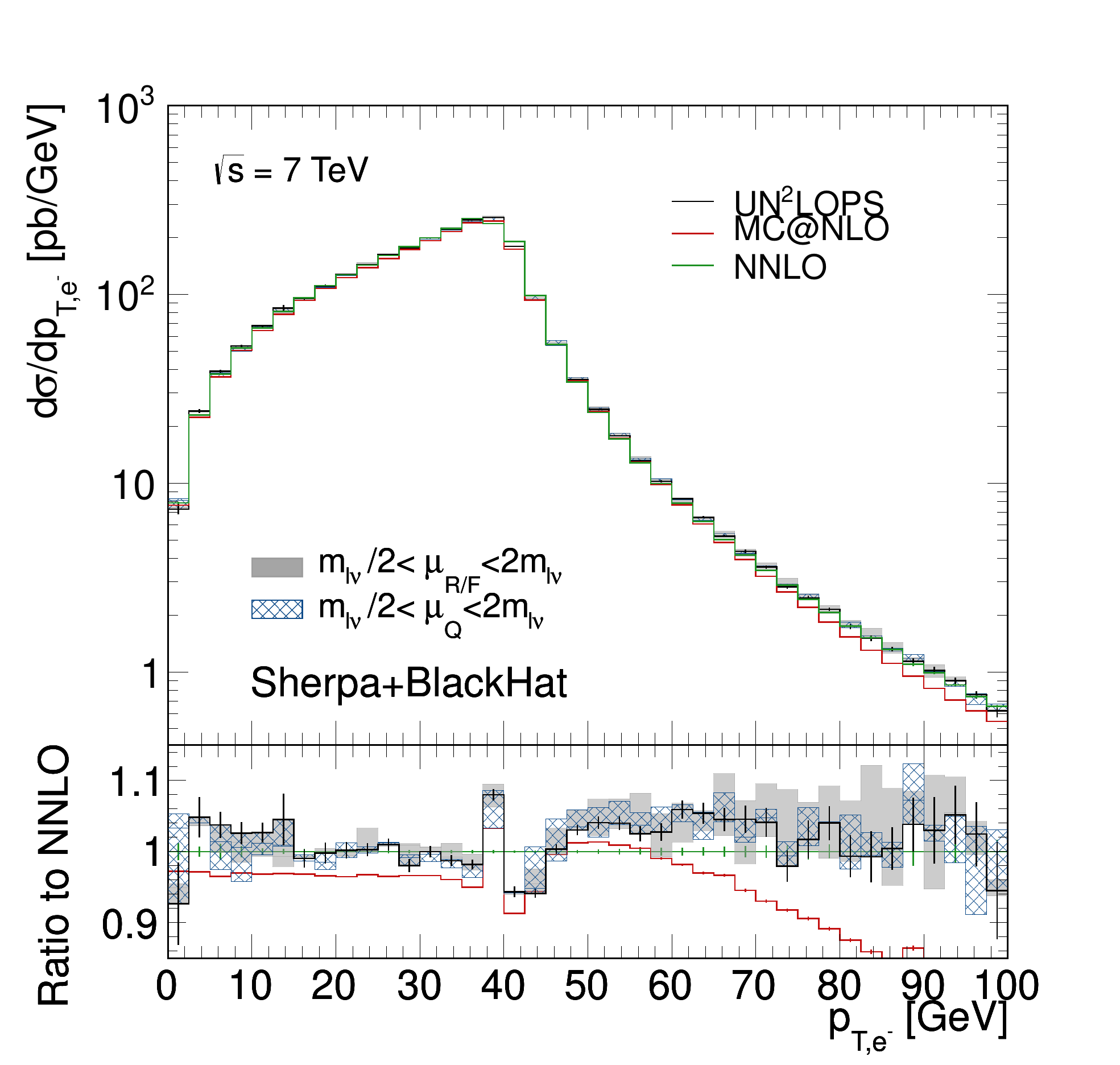}
\includegraphics[width=\halfplotwidth\textwidth]{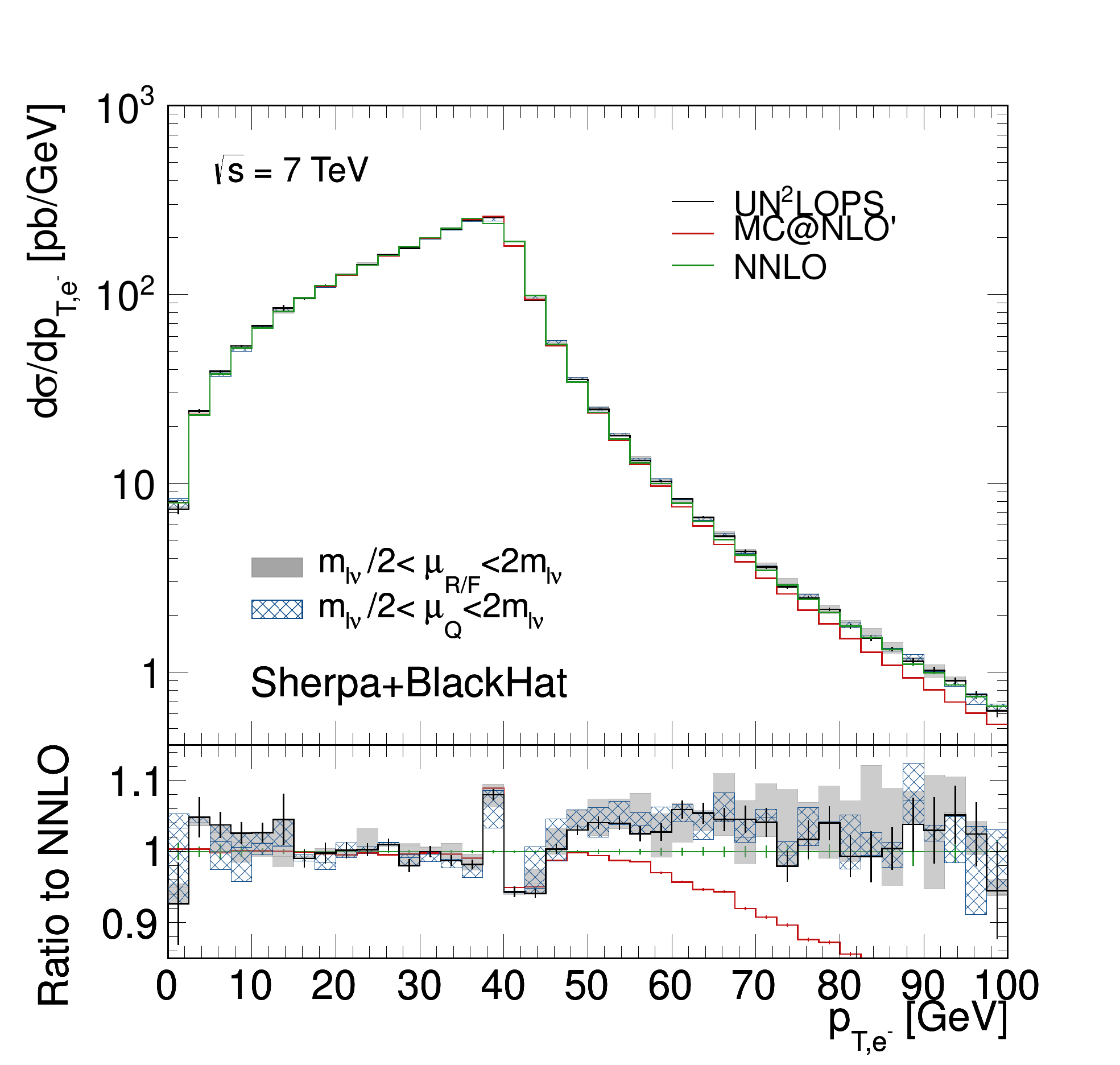}

\caption{Charged current Drell-Yan lepton pair production, for two different PDF choices. 
{\it Upper left}: Pseudorapidity of the positron at NLO and NNLO accuracy. NLO PDFs used in the NLO calculation. 
{\it Upper right}: Pseudorapidity of the positron. NNLO PDFs used in the NLO calculation. 
{\it Lower left}: $p_\perp$ of the positron. NLO PDFs used in \MCatNLO. 
{\it Lower right}: $p_\perp$ of the positron. NNLO PDFs used in \MCatNLO. 
}
\label{fig:pdf_secret_w}
\end{figure*}

Figure \ref{fig:pdf_secret_w} highlights an 
interesting feature of the NNLO corrections to neutral and charged current
Drell-Yan lepton pair production. For inclusive observables,
using a NNLO PDF for a NLO calculation reproduces the full NNLO
calculation very well, both in normalization and in shape. This is clearly
a very process-dependent statement, and it breaks down once an observable
depends not only on the Born degrees of freedom, as shown in the lower right
panel of Figure \ref{fig:pdf_secret_w}: In the phase space region
which can only be accessed by giving the lepton-pair system transverse 
momentum ($p_T>40$~GeV), the NNLO result cannot be mimicked by a NLO calculation.
In this region the improvement obtained from \UNNLOPS is apparent.

\begin{figure*}[t]
\centering
    \includegraphics[width=\halfplotwidth\textwidth]{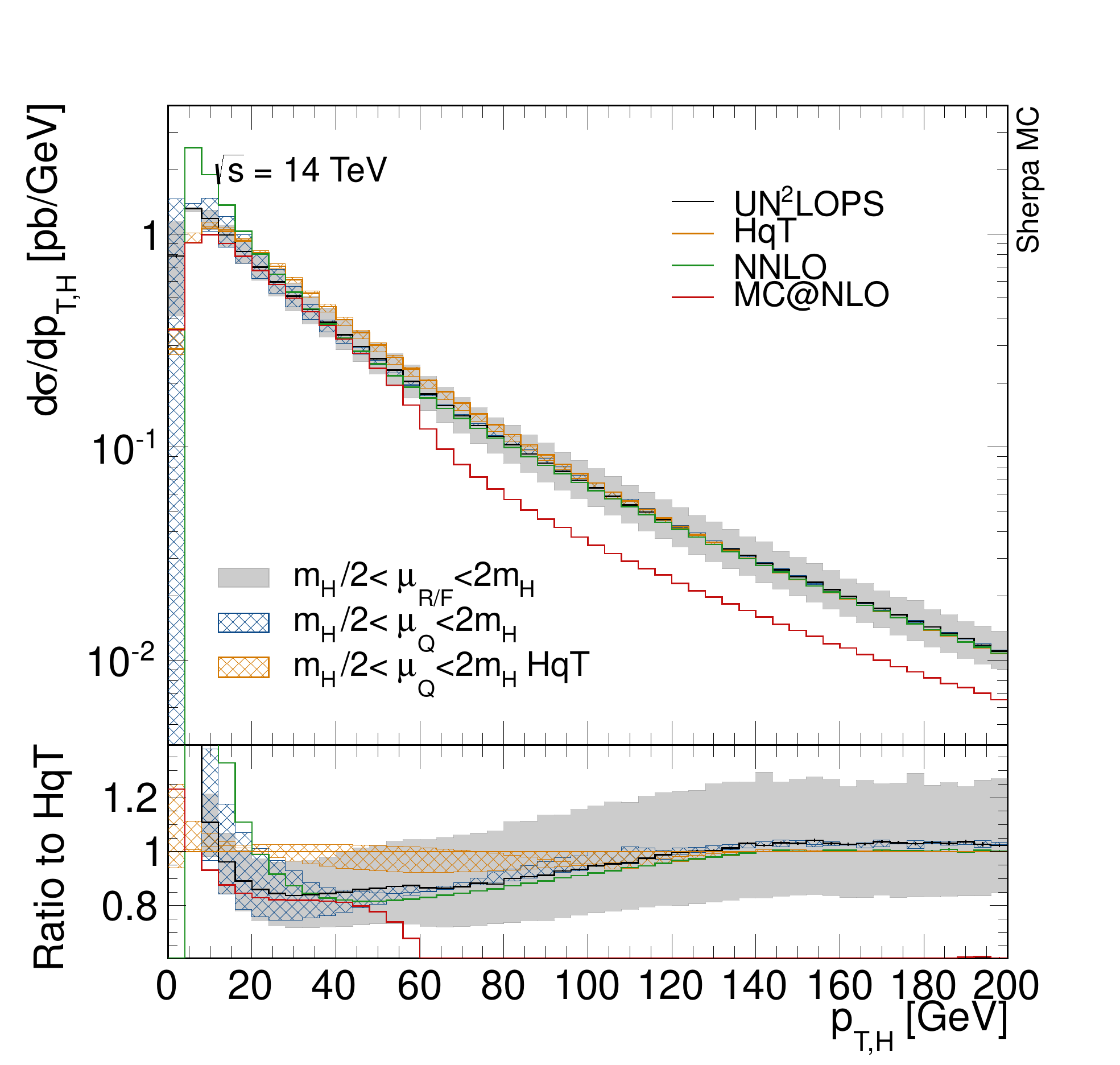}
    \includegraphics[width=\halfplotwidth\textwidth]{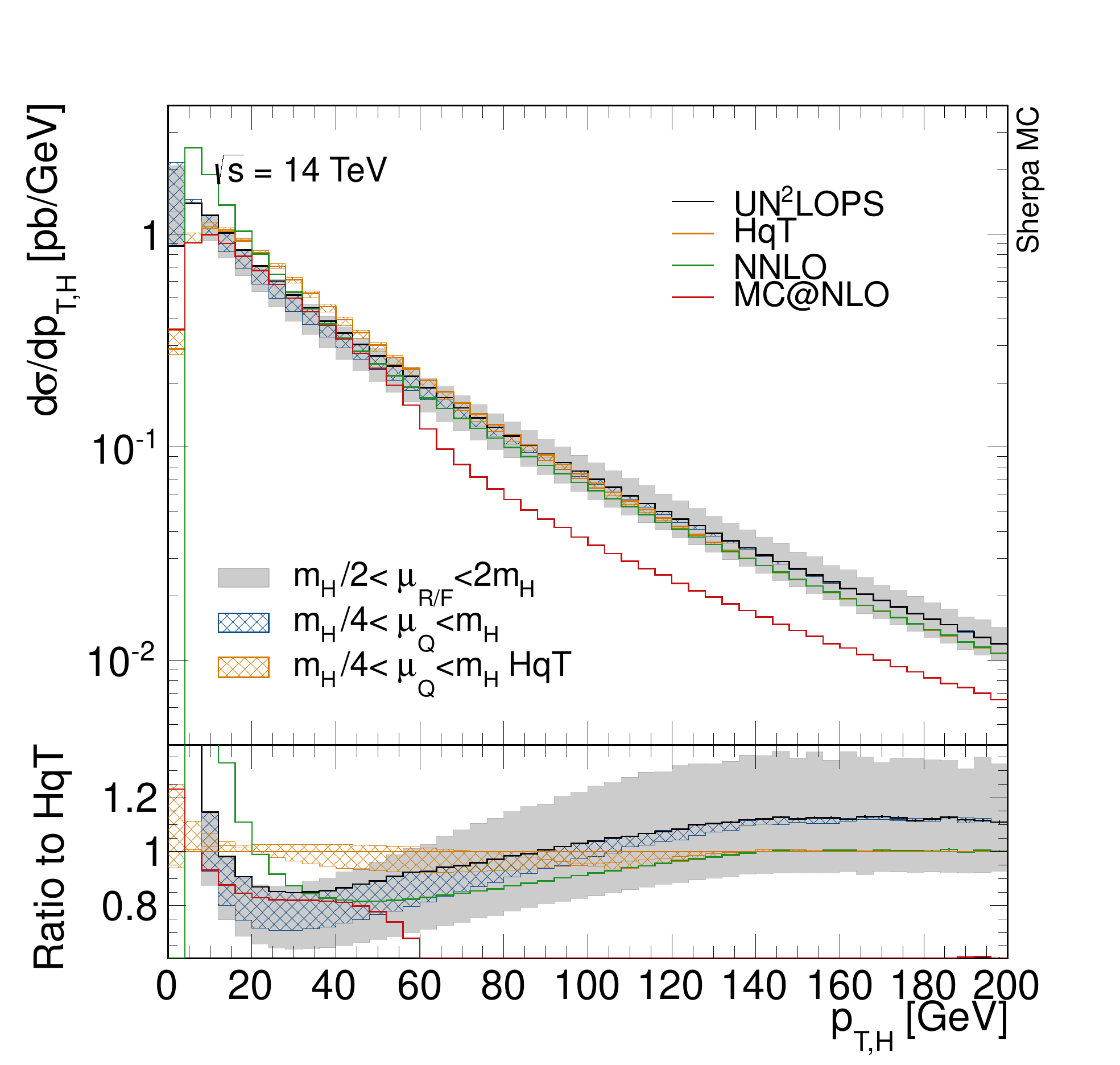}
\caption{Higgs boson $p_\perp$ spectrum in
    individual matching (left) and factorized matching (right).
}
\label{fig:higgs_matching}
\end{figure*}

The \UNNLOPS prescription has also been applied to ~\cite{Hoche:2014dla}. 
Figure \ref{fig:higgs_matching} exemplifies the residual uncertainties of the 
NNLO matched calculation in Higgs-boson production through gluon fusion. We use 
two different ways to include the Wilson coefficient for the $ggh$ vertex~\cite{Hoche:2014dla}:
A factorized matching scheme which is reminiscent of the \POWHEG strategy, and an individual 
matching scheme that somewhat mimics the \MCatNLO procedure. 
The results are as expected: The factorized approach leads to a harder tail in 
the $q_\perp$ distribution, whereas the individual matching has a softer tail 
and a small enhancement for medium $q_\perp$ values. The individual matching shows
better agreement with the NNLO+NNLL result of the HqT program~\cite{Bozzi:2003jy}. The
uncertainty due to varying the parton shower starting scale becomes appreciable
for small $q_\perp$ values, and is significantly larger than the resummation
scale variation in HqT. This might be taken as indication that a more accurate
parton shower would be beneficial.

\section{Conclusions}

We have reviewed the current status of matching and merging parton
shower resummation and fixed order calculations. Some state-of-the-art 
NLO merging methods have recently been molded into NNLO matching methods. 
The prerequisite for these extensions was a well-defined one-jet cross section, 
which was then updated to NNLO accuracy for the inclusive process. Results
of the \UNNLOPS scheme as implemented in \Sherpa have been presented. This 
implementation includes new NNLO fixed-order calculations for (neutral and 
charged current) Drell-Yan lepton pair and (gluon-fusion initiated) Higgs-boson 
production. When applied to the Drell-Yan process, we find that the NLO results, 
when computed with NNLO PDFs, reproduce the full NNLO results for inclusive observables.
For Higgs-boson production at NNLO+PS accuracy, two schemes were 
presented, highlighting some residual uncertainties of the matching.

Work supported by the US Department of Energy under contract DE--AC02--76SF00515.



\end{document}